\documentclass[12pt,preprint,sort&compress]{elsarticle}
\journal{Carbon}
\pdfoutput=1
\makeatletter
\def\ps@pprintTitle{%
 \let\@oddhead\@empty
 \let\@evenhead\@empty
 \def\@oddfoot{}%
 \let\@evenfoot\@oddfoot}
\makeatother

\usepackage{graphicx}  
\usepackage{dcolumn}   
\usepackage{bm}        
\usepackage{amssymb}   

\usepackage{setspace}
\usepackage{amsmath} 
\usepackage{amsthm} 

\usepackage{multicol}
\usepackage{fullpage}

\usepackage[english]{babel}   

\usepackage{tabularht} 
\usepackage{booktabs} 
\usepackage{appendix} 

\begin{document}
\begin{frontmatter}
	
\title{Structure and dynamics investigations of a partially hydrogenated graphene/Ni(111) surface}
	
	\author[ILL,CAM]{Emanuel Bahn\corref{cor}}\ead{em.bahn@gmail.com}
	
	\author[CAM,GRZ]{Anton Tamt\"ogl\corref{cor}}\ead{tamtoegl@gmail.com}
	
	\author[CAM]{John Ellis}
	
	\author[CAM]{William Allison}
	
	\author[ILL]{Peter Fouquet\corref{cor}}\ead{fouquet@ill.eu}
	
	\cortext[cor]{Corresponding author}
	\address[ILL]{Institut Laue-Langevin, 71 Avenue des Martyrs, 38042 Grenoble CEDEX 9, France}
	\address[CAM]{Cavendish Laboratory, 19 J J Thomson Avenue, Cambridge CB3 0HE, United Kingdom}
	\address[GRZ]{Institute of Experimental Physics, Graz University of Technology, Petersgasse 16, 8010 Graz, Austria}

\begin{abstract}
Using helium-3 atom scattering, we have studied the adsorption kinetics, the structure and the diffusional dynamics of atomic hydrogen on the surface of a graphene monolayer on Ni(111). Diffraction measurements reveal a 4$^\circ$ rotated rectangular hydrogen overstructure. Hydrogen adsorption and desorption exhibit activation barriers of $E_a=(89\pm7)$~meV and $E_d=(1.8\pm0.2)$~eV, respectively. Helium-3 spin-echo measurements showed no decay of the spatial correlation function (or intermediate scattering function) within the time range of the spectrometer. Hence, we are able to set lower limits for a possible hydrogen surface diffusion rate.
\end{abstract}

\begin{keyword}
	Graphene \sep
	Ni(111) \sep
	Hydrogenation\sep
	Helium Atom Scattering
\end{keyword}

\end{frontmatter}

\section{Introduction}
\label{sec:intro}

The existence of an ordered hydrogenation of graphene remains an open question to date although it is of fundamental interest for the tuning of graphene properties \cite{Lin:2015gq}. The existence of a stable fully hydrogenated graphene \cite{Novoselov:2004ub,Geim:2007ue} layer (graphane) was theoretically predicted almost a decade ago \cite{Sofo:2007bp}, but the synthesis of an almost perfect graphane layer has only recently been reported \cite{Yang:2016wb}. A review on graphane and hydrogenated graphene has been published by Pumera \textit{et al.} \cite{Pumera:2013gj}.
In perfect graphane, one hydrogen atom binds to each carbon atom in an sp$^3$ hybridised state in an alternating manner, with the hydrogen atom being located either above or below the 2D plane. This leads to four different possible conformations: the chair, boat, twisted-chair, and twisted-boat conformations \cite{Pumera:2013gj}.

The first successful hydrogenation of graphene was reported by Elias \textit{et al.} \cite{Elias:2009gh}, where free-standing graphene and graphene on a SiO$_2$ substrate were used. On top of a substrate, only single-sided hydrogenation of graphene is possible, which has been predicted to create a disordered material \cite{Elias:2009gh}. A graphene layer that is fully hydrogenated on one side is being referred to as graphone. Graphone has recently been successfully synthesised from graphene/Ni(111) in a reversible manner \cite{Zhao:2015fu}.
The partial one-sided hydrogenation of a graphene sheet has also recently been achieved by splitting of intercalated water in the graphene/Ni(111) system at room temperature \cite{Politano:2016jb}.

Before the theoretical discovery of graphane and graphone, the hydrogenation of the graphite(0001) surface had been studied intensely for various interests, such as the mechanism of H$_2$ formation in the interstellar medium. A disordered formation of hydrogen dimers on graphite was observed at low hydrogen coverage \cite{Hornekaer:2006bc,Hornekaer:2006iw}, while at high coverage the appearance of a triangular structure was observed that occupies only one of the two trigonal sublattices, into which the (0001) surface can be separated \cite{Hornekaer:2007kc}.
This preferential sublattice adsorption is also found in the hydrogenation of metal supported graphene, where the underlying metal lattice often plays an important role.

Ng \textit{et al.} have shown that on the weakly interacting metal-graphene systems such as graphene/Ir(111) and graphene/Pt(111) H atoms form a graphone-like structure in the valleys of the moir\'e pattern \cite{Ng:2010jg} which is caused by a lattice mismatch between the graphene and the metal lattice. On the strongly bound and lattice matched graphene/Ni(111) system, however, a blocking of not only one sublattice, but also of nearest neighbour adsorption sites of the same sublattice has been observed \cite{Ng:2010jg}. This is in good agreement with our findings, as will be discussed later. On the other hand, on the same graphene/Ni(111) system, a full hydrogenation of one sublattice, and therefore the synthesis of graphone was found in a different study \cite{Zhao:2015fu}. The major difference between the two studies is that in the latter one, hydrogenation took place at a much lower temperature (170~K compared to 300~K).

On graphene/SiC, dimer formation at low coverage and disordered cluster formation at high coverage have been observed. In contrast to the graphite(0001) surface, hydrogen monomers were also found on graphene/SiC, suggesting a stronger binding to the surface in this case \cite{Guisinger:2009bm,Balog:2009bn}. On graphene/Cu(111), a structural arrangement has been observed by scanning tunnelling microscopy (STM) \cite{Lin:2015gq}, where three different configurations were found, all with a preference to sublattice adsorption. Such a sublattice configuration is predicted to cause ferromagnetism in graphene \cite{Zhou:2009ck}. These effects are linked to the fact that the adsorption of an H atom on a graphene surface creates a distortion in both electron and spin density that exhibits a ($\sqrt{3}$x$\sqrt{3}$)\textbf{R}30$^\circ$ overstructure and that extends over several nm, as shown by ab-initio calculations and STM/ atomic-force microscopy (AFM) measurements \cite{Ruffieux:2000wb,Yazyev:2007hp,Casolo:2009du}. 

Directly connected to the question of an ordered hydrogenation is the question if a diffusion of H adatoms on the graphene/Ni(111) surface is in general possible (i.e., if the diffusion barrier is lower than the desorption barrier) and, if possible, at what rates it would occur. On the graphene/Ni(111) surface, density-functional theory (DFT) calculations predict a diffusion barrier of 1.9~eV, which is lower than the desorption barrier for a single H atom of 2.25~eV \cite{Zhao:2015fu}. Quantum transition-state theory (TST) calculations predict H atom diffusion on a free-standing graphene surface with a diffusion barrier of 0.71 eV \cite{Herrero:2009vn}. DFT calculations predict a diffusion barrier of 1.25~eV for single H atoms, but only 0.46~eV for an H atom in the vicinity of a second atom \cite{Ferro:2003vl}. On a graphene bilayer, Kinetic Monte Carlo (KMC) calculations predict dimer formation via diffusion and subsequent desorption within a few minutes \cite{Moaied:2015di}.

Here, we present our findings from helium atom scattering (HAS). Various pristine graphene/metal systems have been studied successfully with HAS previously, where precise values for the surface Debye temperature, electronic corrugation, or possible moir\'e patterns could be determined 
\cite{Borca:2010ib,Politano:2011jw,Gibson:2014gf,Politano:2015jd,Shichibe:2015fp,AlTaleb:2015bx,Maccariello:2016de}. Notably, neutral He atom beams with energies in the order of 5\,--\,10~meV are perfectly suited to probe H overlayers in an inert, completely non-destructive manner. HAS provides, furthermore, a precise probe for both adsorbate coverage and adsorbate structure due to the very large scattering cross section of surface defects and adsorbed molecules \cite{Farias:1998td}. This allowed us to study both, hydrogen sorption kinetics and structural adsorbate ordering. In addition, the spin-echo technique allowed us to gain information on surface dynamics in a time range from sub-picoseconds to about one nanosecond. We made use of this to exclude a diffusion of H atoms over a large time scale.

\section{Experimental Details}
\label{sec:experiment}
	
\subsection{Hydrogenation of graphene/Ni(111)}

All measurements have been performed on the Cambridge helium-3 spin-echo spectrometer (HeSE) \cite{Jardine:2009gj,Fouquet:2005ki}. We have published the characterisation and growth of the graphene layer on a Ni(111) surface elsewhere \cite{Tamtogl:2015jb}. Briefly, the nickel (Ni) (111) single crystal used in the study was a disc with a diameter of 10~mm and a thickness of 1~mm. The crystal was mounted on a sample holder, which can be heated using radiative heating from a filament on the backside of the crystal or cooled down to 100~K using liquid nitrogen or 45~K using liquid helium, respectively. The sample temperature was measured using a chromel-alumel thermocouple. Prior to the measurements, the surface was cleaned by Ar$^+$ sputtering and annealing at 870~K. A monolayer of graphene on Ni(111) was grown by dosing ethene (C$_2$H$_4$) while heating the Ni crystal (730~K) over several hours. 

In a series of experiments, the graphene/Ni(111) sample was heated to different temperatures and hydrogen gas was injected into the vacuum chamber using a microcapillary array beam doser at a distance of 50~mm to the sample. A cracking filament (4~A, 1~V DC) was used to produce atomic hydrogen. 
A mass spectrometer was routinely used during all measurements to exclude possible contaminations of the chamber.
	
We studied the hydrogen adsorption kinetics by monitoring the attenuation of the specular helium-scattering signal during hydrogenation of the graphene surface. Adsorption was monitored at 400\,--\,700~K at a dosing pressure at the surface of $3\cdot 10^{-6}$~mbar. Subsequently, isothermal desorption was monitored at different temperatures using the same approach.
	
\subsection{Diffraction and spin-echo measurements}
	
After a maximum possible hydrogenation of the graphene surface (confirmed by a maximum attenuation of the specular scattering signal), helium diffraction studies were performed in the fixed 44.4$^\circ$ source -- detector scattering geometry of the HeSE apparatus. The surface temperature of 400 K was chosen because a contamination through intercalation at defect sites cannot be ruled out at room temperature, although the graphene/Ni(111) surface is highly inert.
In the case of HAS, the momentum transfer $\Delta \mathbf{k}$ can be separated into a component parallel to the surface and perpendicular to the surface (designated by subscript z): $\Delta \mathbf{k} = ( \Delta \mathbf{K}, \Delta k_{z} )$. For a diffraction scan the momentum transfer parallel to the surface, given by $\lvert \Delta \mathbf{K} \rvert = \lvert \mathbf{k_i} \rvert ( \sin (\gamma_f ) - \sin (\gamma_i ) )$, is varied by changing the incident angle $\gamma_i$. Here, $\mathbf{k_i}$ is the incident wave vector and $\gamma_i$ and $\gamma_f$ are the incident and final angle with respect to the surface normal, respectively.
	
Helium spin-echo measurements were subsequently performed in the same set-up at temperatures of 400, 500 and 600~K, respectively. At 500 and 600~K, a constant H overpressure was applied to ensure a constant surface coverage despite desorption. In spin-echo spectroscopy the intermediate scattering function (ISF), $I(\Delta K,t)$, is directly measured in form of a polarisation amplitude \cite{Mezei:1972wv}. Structural reconfigurations of the surface from e.g. adsorbate diffusion cause a decay of $I(\Delta K,t)$ over the spin-echo time, $t$. For simple diffusional processes this decay bears the shape of an exponential decay \cite{Alexandrowicz:2007kd}. 

\section{Results and Discussion}
\label{sec:results}

\subsection{Isothermal Hydrogen Adsorption}

Adsorption measurements were performed by monitoring the specularly reflected helium signal in real time while exposing the graphene surface to a constant partial pressure of atomic hydrogen, that had been dissociated by a cracking filament. This was repeated at different surface temperatures in the range 400 -- 700~K. Adsorption was observed up to temperatures below 600~K, above which an adsorption-desorption equilibrium was rapidly reached and thus no substantial scattering-signal attenuation occurred.

\begin{figure}[htb]
	\centering	
	\includegraphics[width=0.5\linewidth]{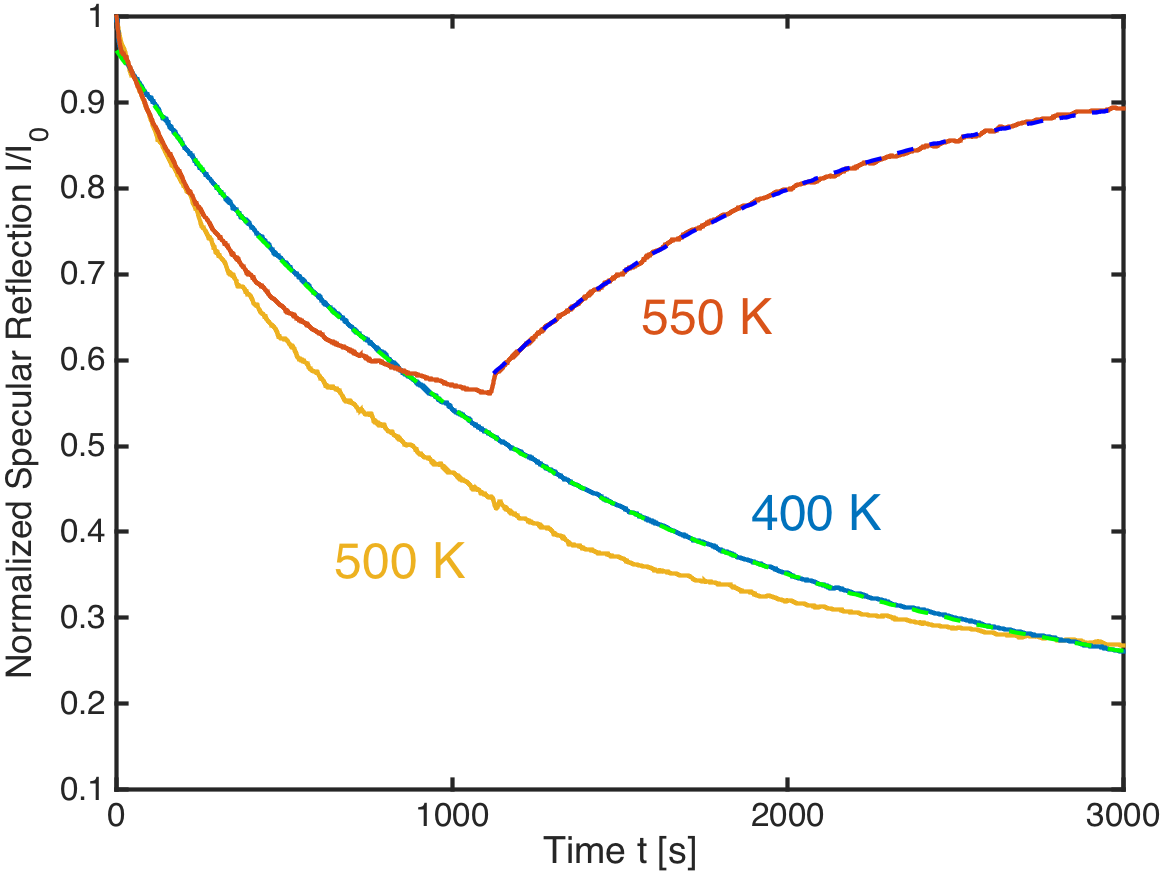}
	\caption{(Colour online) Normalised specular helium-scattering signal $I/I_0$ as a function of time during exposure of the graphene/Ni(111) surface to atomic hydrogen in the temperature range 400\,--\,550~K. Green dashed line: exponential decay fitted to the scattering signal at 400~K during adsorption; blue dashed line: exponential recovery fitted to the scattering signal at 550~K during desorption (the hydrogen overpressure was turned off after about 1100 s). Instant jumps of the scattering signal appear at the start of dosing (0 s) and after the dosing has been stopped (1100 s) due to attenuation of the helium beam from ambient hydrogen gas.}
	\label{fig:uptake}
\end{figure}

Fig. \ref{fig:uptake} shows the specularly reflected helium-scattering signal during adsorption of atomic hydrogen on the graphene/Ni(111) surface for three different temperatures: 400~K, 500~K, and~550 K. In this range, a single exponential decay of the scattering signal was observed, as shown for the 400~K data by a dashed green line. The perfect exponential behaviour indicates a lack of reorientation of adsorbed atoms due to attractive or repulsive interaction with other adatoms, resulting in a slower-than-exponential or a faster-than-exponential decay, respectively \cite{Comsa:1989ua}.

In contrast to the high temperature adsorption of atomic hydrogen, dosing molecular hydrogen through a microcapillary array beam doser at surface temperatures above 100~K does not give rise to any changes of the specularly reflected He signal. This indicates the total lack of adsorbed molecular hydrogen on the graphene/Ni(111) surface. Only upon cooling the surface to temperatures between 45\,--\,80~K a significant change in the specular helium reflection was observed. In subsequent diffraction scans no additional diffraction peaks were observed for molecular hydrogen adsorbed on graphene/Ni(111). Hence we can exclude the adsorption of molecular hydrogen at temperatures above 100~K. The graphene/Ni(111) surface seems to be largely inert with respect to the adsorption of molecular hydrogen and H$_2$ needs to be pre-dissociated before adsorption at higher temperatures occurs.

\begin{figure}[htb]
	\centering	
	\includegraphics[width=0.5\linewidth]{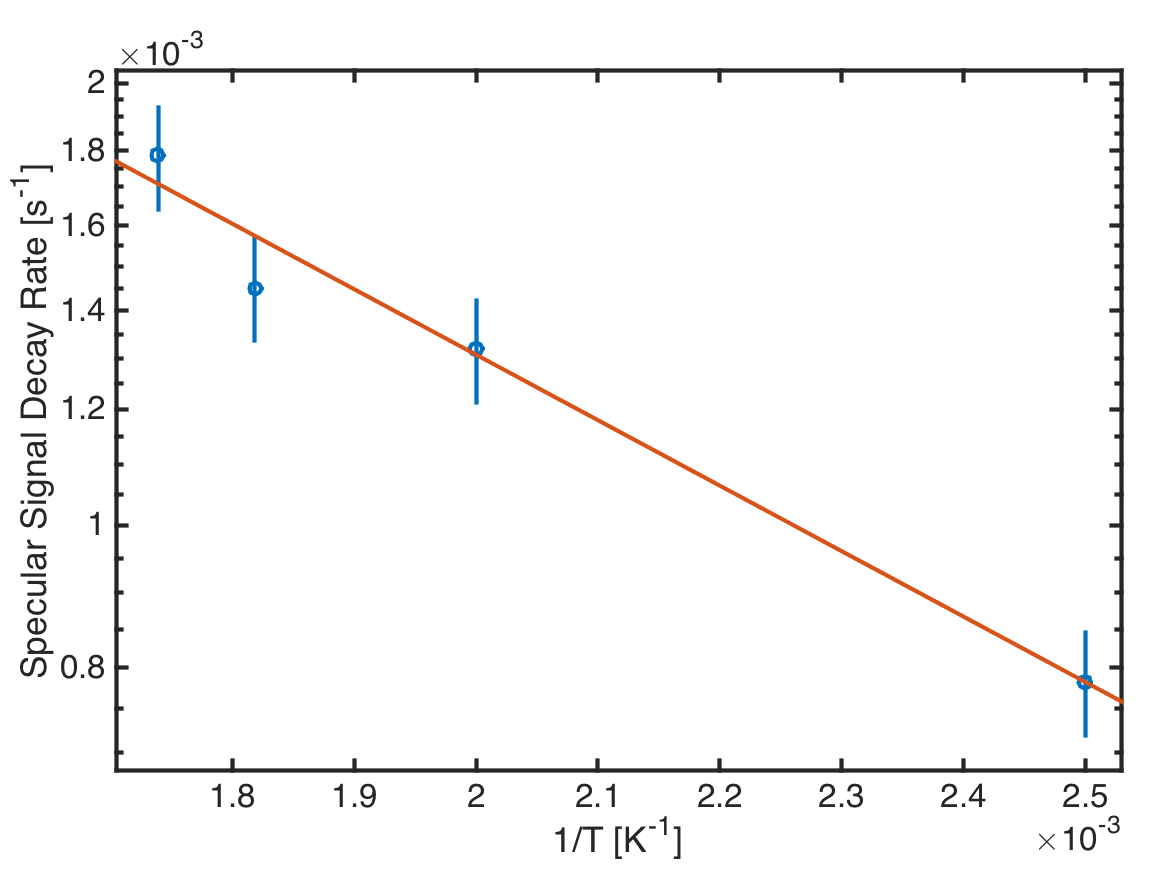}
	\caption{Arrhenius plot of the specular scattering-signal decay rate during isothermal hydrogen adsorption versus inverse temperature. The slope of the linear fit returns an adsorption barrier $E_a=(89\pm7)$~meV.}
	\label{fig:uptake_rate}
\end{figure}

Since we have no absolute calibration of the cracking filament efficiency, we cannot deduce the flux of atomic hydrogen to the surface from the monitored H$_2$ background pressure. We therefore have to rely on existing information on the helium scattering cross section from single H atoms in order to relate the specularly reflected signal to a surface coverage. The helium scattering cross section $\Sigma$ of H on transition metal and semimetal surfaces is typically around $10 - 12~\mbox{\AA}^2$ \cite{Poelsema:1981ua,Comsa:1989ua,Kraus:2013eu}. 
Since $\Sigma$ is comparable to the graphene/Ni(111) unit cell size ($5.4$~\AA$^2$), we expect the effect of overlap to be negligible. Indeed, all measured H adsorption isotherms exhibit the shape of an exponential decay over time at constant partial H pressure, as would be expected from the Langmuir adsorption model where adsorption is limited to one monolayer, all adsorption sites are equivalent and only one adparticle can reside in an adsorption site. We observed the highest attenuation of about 80~\% at 400~K, where we expect desorption to be negligible. Assuming an approximative scattering cross section of 11~\AA$^2$, this would correspond to one H atom per 2--3 graphene unit cells or an H/C ratio of about 20~\%. The fact that the adsorption curves at higher temperature (500 K and 550 K) exhibit a slightly steeper initial slope with respect to the curve at 400 K indicates that there exists a small activation energy for the adsorption of atomic hydrogen. Fig. \ref{fig:uptake_rate} shows the rates of exponential decay of the specularly reflected signal during adsorption as a function of the inverse temperature in an Arrhenius plot. The errorbars indicate an approximative relative statistical error of 8~\% which was obtained by comparing repeated measurements. We expect the source of this error to lie in a gradual misalignment of the sample with respect to specular and/or in a fluctuating efficiency of the hydrogen cracking filament. The linear dependence indicates an activation barrier to adsorption. From the slope of a fitted straight line we can deduce an adsorption barrier, $E_a=(89\pm7)$~meV.
DFT calculations on free-standing graphene and graphite predict an adsorption barrier for a single H atom of 0.14~eV \cite{Casolo:2009du}, 0.2~eV \cite{Jeloaica:1999tk,Sha:2002tc}, 0.25~eV \cite{Sljivancanin:2009dl} and a reduced barrier down to 0~eV, when the adsorption happens in the neighbourhood of an adsorbed H atom and on the same sublattice, since local magnetization reduces the adsorption barrier \cite{Sljivancanin:2009dl,Casolo:2009du}. Our observation of a relatively low adsorption barrier might therefore hint towards a preferential sublattice adsorption.

We subsequently monitored isothermal desorption from the surface by monitoring the recovery of the specular helium-scattering signal from a covered surface at different constant surface temperatures. In the range 500 -- 600~K, rapid desorption from the covered surface was observed in the form of an exponential recovery of the scattering-signal. Fig. \ref{fig:uptake} shows the specular scattering signal recovery during isothermal desorption at 550~K (hydrogen dosing was switched off after $\simeq1100$ s) together with a fitted exponential curve, shown by a blue dashed line. Due to gradual misalignment of the sample away from the specular position over time (which stems from thermal expansion of the sample environment at elevated temperatures), the specular signal does not recover entirely to its initial value. Re-aligning the crystal to the specular reflection condition after the measurement restored the original intensity of the clean surface. Since this behaviour does not cause significant changes of the observed decay rates we did not apply any corrections.

\begin{figure}[htb]
	\centering
	\includegraphics[width=0.5\linewidth]{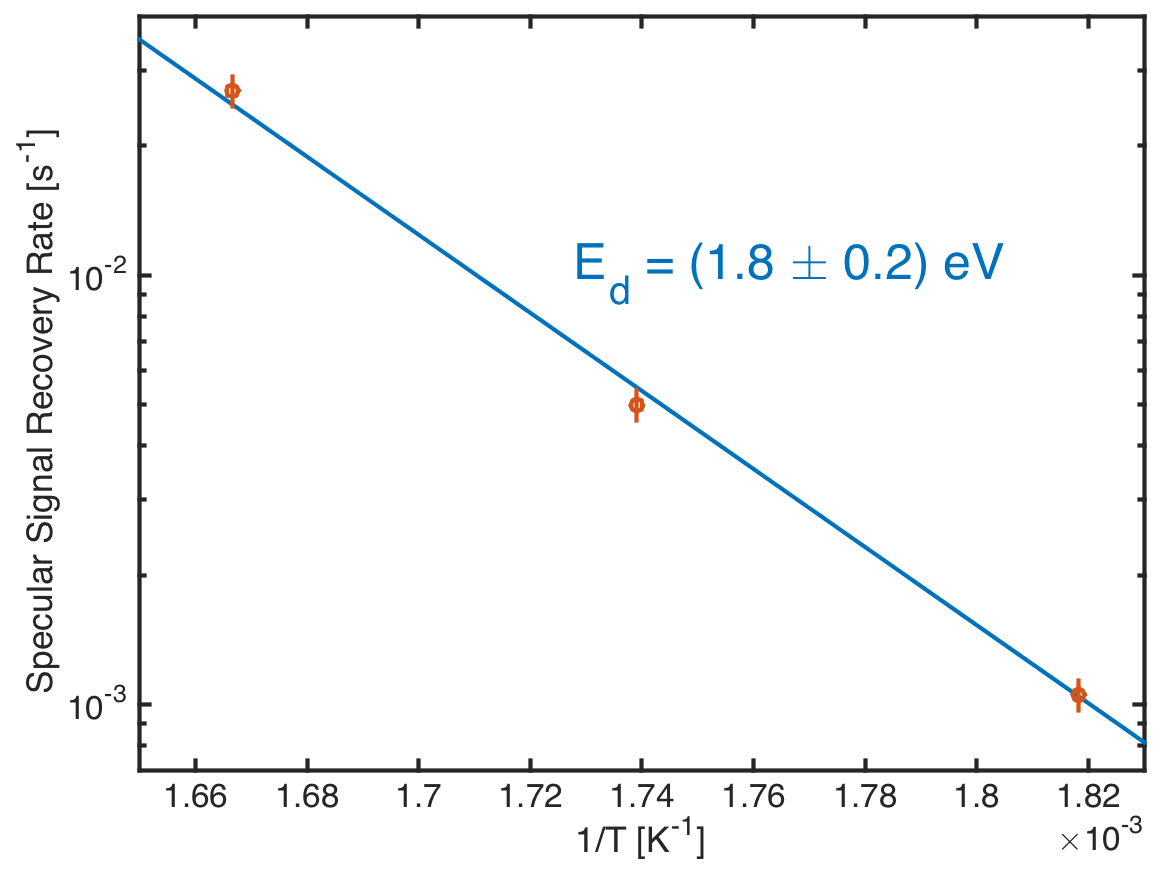}
	\caption{Arrhenius plot of the specular scattering-signal recovery rate during isothermal hydrogen desorption versus inverse temperature. Errorbars indicate the standard error obtained from the exponential fits of the isothermal decays. The straight line shows a linear fit to the data points. The slope of the linear fit returns an activation energy of desorption $E_d=(1.8\pm0.2)$~meV.}
	\label{fig:recoveryrate}
\end{figure}

With respect to conventional temperature programmed desorption (TPD) this method has the advantage that it is very sensitive to small changes of the hydrogen coverage due to the large scattering cross section of single adsorbates on a surface. Conventional TPD is often limited to studies of D$_2$ adsorption/desorption since there is typically always a rather high amount of H$_2$ in the residual gas and the mass of D$_2$ allows a better distinction from the background.
Fig. \ref{fig:recoveryrate} shows an Arrhenius plot of the isothermal hydrogen desorption rates, obtained by fitting an exponential decay to the specular signal recovery during isothermal desorption. Down to a temperature of about 550~K, a straight line can be fitted to the data, which corresponds to activated behaviour for desorption. From this we obtain an activation energy for desorption $E_d=(1.8\pm0.2)$~eV.
This agrees well with results from TPD measurements from a hydrogenated graphene/Ni(111) surface, where two desorption processes with 1.0~eV and 1.8~eV desorption barriers were found \cite{Zhao:2015fu}. Since the lower energy process would lead to rapid desorption already at room temperature, it is probable that we observed only the high-barrier desorption. The authors argue with the support of DFT calculations that the desorption with a barrier of about 1.8~eV is mediated by diffusion towards a meta-dimer as a rate-limiting step, while at higher coverage, diffusion is no longer necessary.
TPD spectra from highly covered graphite surfaces found conflicting values for the desorption barrier of hydrogen with 0.6~eV \cite{Zecho:2002cz} and 1.3--1.6~eV \cite{Hornekaer:2006bc}, respectively. On the other hand, based on DFT calculations, adsorption energies of 1.5-1.7~eV where found, depending on the number of graphene layers \cite{Boukhvalov:2008de}, which agrees very well with our experimental findings. Note however, that precaution must be taken when comparing values calculated for free-standing graphene with measurements on graphene/Ni(111) due to the relatively strong binding of graphene to the Ni substrate.

Moreover, DFT calculations predict a barrier of 1.25~eV for diffusion of a single H atom and 2.8~eV for recombination of two neighbouring H atoms \cite{Ferro:2003vl}. Under ultra-high vacuum conditions, recombinative desorption via the Eley-Rideal mechanism is not possible and it is more likely that individual hydrogen atoms recombine to desorb as molecular hydrogen. Therefore, the question about a possible surface diffusion leading to recombinative desorption arises.

\subsection{Structure}

\begin{figure}[htb]
	\centering
	\includegraphics[width=0.5\linewidth]{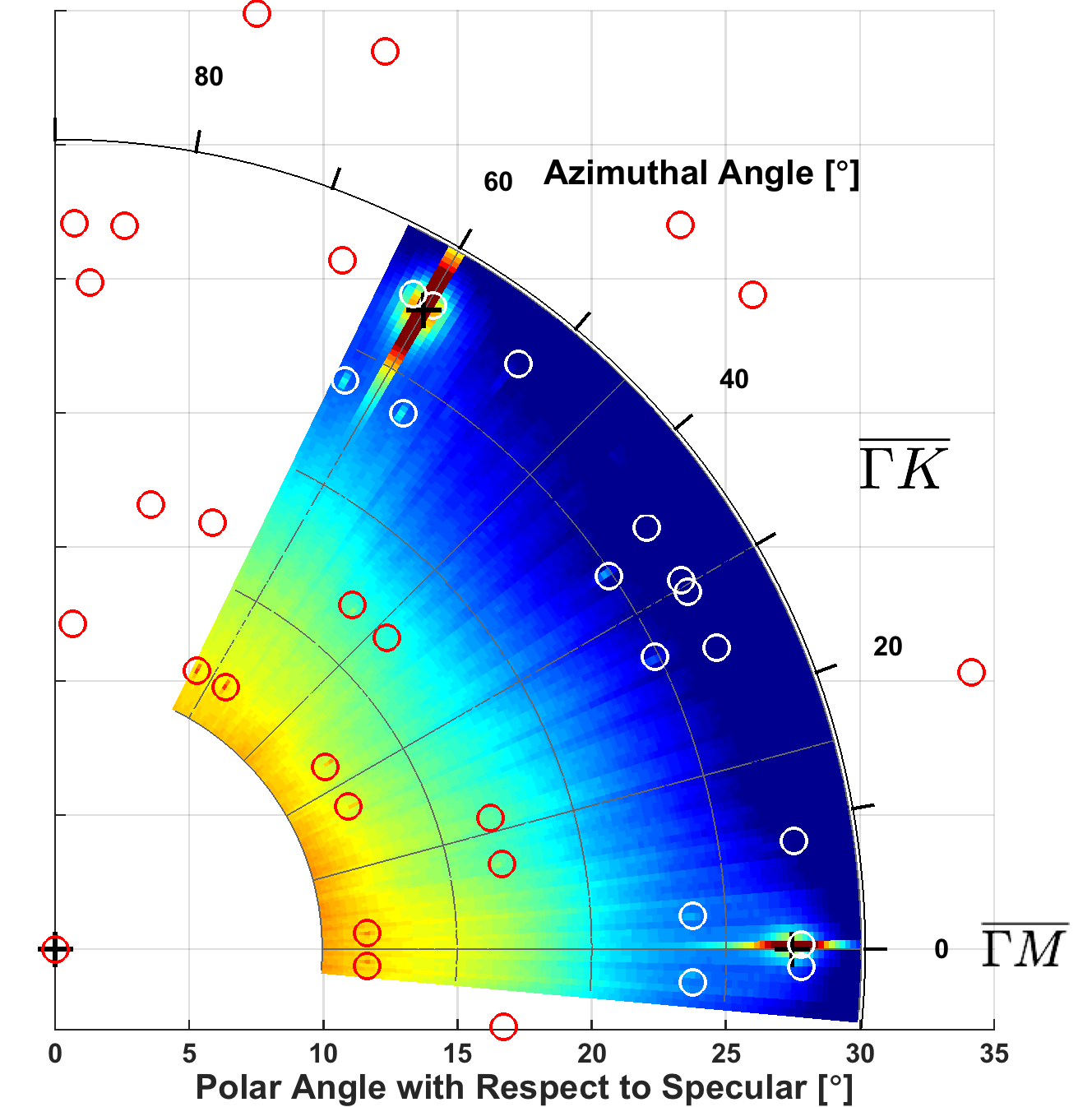}	
	\caption{(Colour online) Two-dimensional colour plot of the helium diffraction signal where warmer colours correspond to a higher scattering intensity. The radial axis corresponds to the angle of incidence, with the specular (mirror) scattering geometry in the centre. The region around the specular position is not shown because of its high intensity. The polar axis corresponds the the azimuthal scattering angle, with $\alpha=0$ corresponding to the $\overline{\Gamma \mathrm{M}}$ azimuth. Black crosses: position of graphene/Ni(111) first order diffraction peaks; red/white circles: theoretical positions of the 4$^\circ$ rotated rectangular overstructure; grey dashed arrows: unit cell vectors of the overstructure. Some of the additional peaks due to the hydrogenation are difficult to spot in the contour plot, however, they are definitely present as can be seen in the one-dimensional scans presented in fig. \ref{fig:diff_1D}.}
	\label{fig:diff_2D}
\end{figure}

We have performed HAS diffraction scans of a fully (up to saturation) hydrogenated graphene/Ni(111) surface along the off-specular angle $\gamma$ and the azimuthal angle $\alpha$. Fig. \ref{fig:diff_2D} shows the diffraction pattern as a polar plot with $\gamma$ along the radial axis and $\alpha$ along the angular component ($\alpha=0$ corresponds to the $\overline{\Gamma \mathrm{M}}$ azimuthal direction). Bright diffraction peaks are visible at the same positions as the first order diffraction peaks of the graphene layer and of the underlying Ni(111) surface. In addition, a large number of smaller diffraction peaks have appeared upon hydrogenation.

\begin{figure}[htb]
	\centering
	\includegraphics[width=0.5\linewidth]{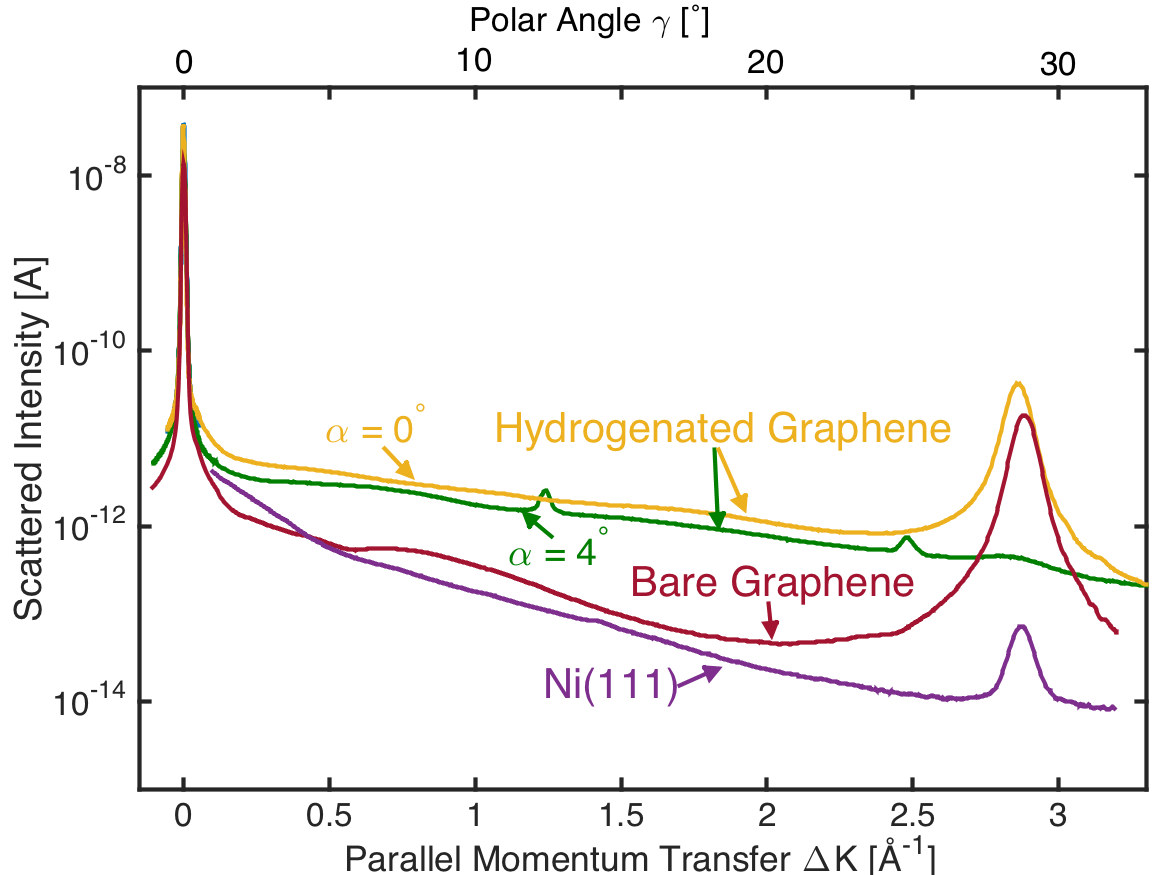}
	\caption{Comparison of the scattered intensities for hydrogenated graphene/Ni(111), bare graphene/Ni(111) and clean Ni(111) versus momentum transfer $\Delta K$. For better orientation, the off-specular polar scattering angle $\gamma$ is indicated on the top axis. All scans were taken at a He beam energy of 8~meV. Graphene/Ni(111) and the pristine Ni(111) surface were scanned at 300~K, the hydrogenated graphene at 400~K.}
	\label{fig:diff_1D}
\end{figure}

Fig. \ref{fig:diff_1D} shows a comparison of diffraction scans along $\gamma$ at the azimuthal angle $\alpha=0^{\circ}$ for the Ni(111) surface, the bare, and the hydrogenated graphene surface. In addition, a diffraction scan through the peaks at $4^\circ$ is shown. For all scans, the scattering signal is plotted against the parallel momentum transfer $\Delta K$ (on the bottom axis). The corresponding off-specular scattering angles $\gamma$ are indicated on the top axis. Note that we did not pursue any corrugation analysis based on close-coupling calculations at this point, since the angular diffraction scans lack bound state resonances which could be used to determine the atom-surface interaction potential \cite{MayrhoferReinhartshuber:2013bg}.

\begin{figure}[htb]
	\centering
	\includegraphics[width=0.5\linewidth]{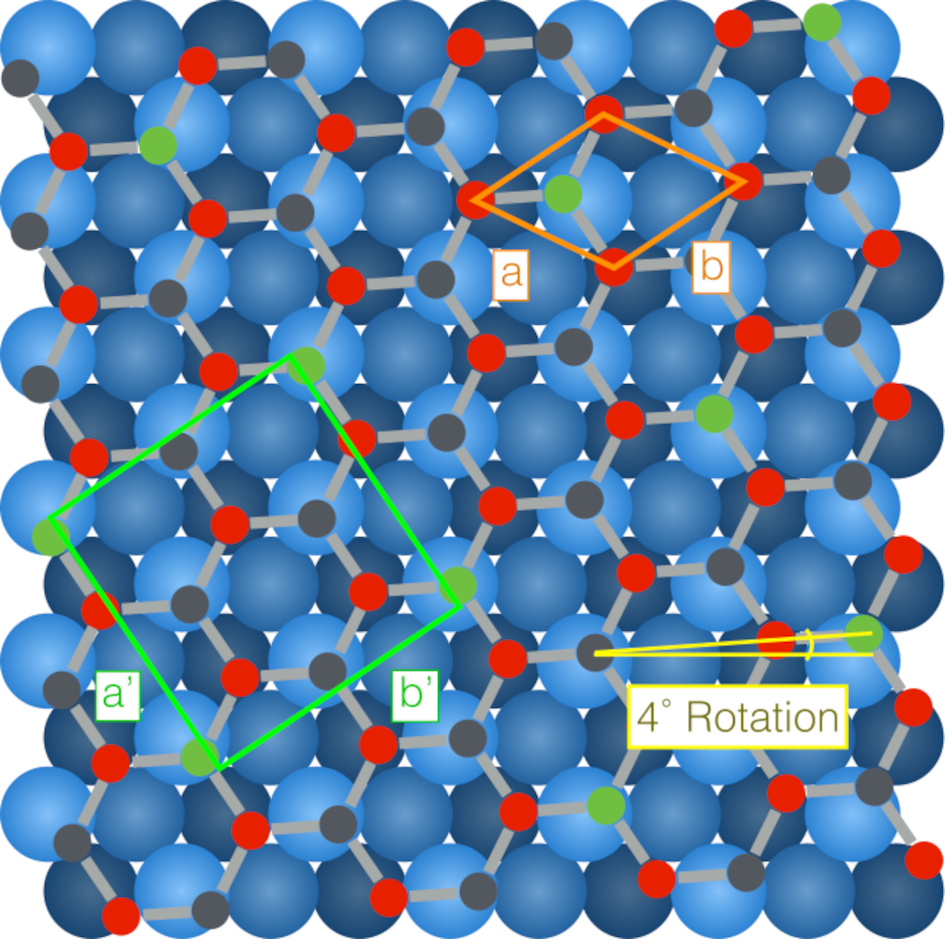}
	\caption{Illustration of the proposed $4^\circ$ rotated structure of hydrogenated graphene/Ni(111). Green rectangle: unit cell of observed diffraction pattern; orange rhombus: graphene/Ni(111) unit cell for comparison; Blue spheres in background: Ni(111) surface; Hexagonal structure: graphene layer, $4^\circ$ rotated but undistorted;  green dots: C Atoms favourable for H  adsorption; dark red dots: fcc sublattice that is energetically unfavourable for H adsorption; grey dots: C atoms of same sublattice as adsorption sites, but energetically unfavourable as next-nearest neighbours of adsorption sites. The relative position of the graphene layer on-top of the Ni(111) surface is chosen arbitrarily since our diffraction measurements allow no conclusions on this matter. Please note that probable distortions of the graphene layer due to hydrogenation have not been taken into account in this illustration.}
	\label{fig:illustration}
\end{figure}

Via the condition of elastic diffractive scattering we could determine the geometry of the hydrogenated graphene surface from the positions of the diffraction peaks: The positions of the bright peaks could either stem from bare graphene or from graphone-like regions, which cause diffraction peaks at the same position \cite{Lin:2015gq}.
The positions of the smaller peaks are well reproduced by different domains of an $\alpha=\pm4^\circ$ rotated rectangular superstructure with $a'=(1.98\pm0.06)\cdot a=(4.93\pm0.15)\,\mbox{\AA}$ and $b'=(1.9\pm0.06)\cdot b=(4.73\pm0.14)\,\mbox{\AA}$ unit cell size ($a=b=2.49$~\AA{} are the moduli of the basis vectors of the graphene/Ni(111) unit cell). This structure is surprising since it is much larger than the boat- and chair-conformation. However, on graphene/Ni(111), photoemission spectroscopy measurements have predicted an adsorption barrier that prevents adsorption on neighbouring C atoms on the "on top" sublattice, resulting in a minimum distance between two adsorbed H atoms of $\sqrt{3}a$, assuming that the H atoms adsorb directly on top of the C atoms \cite{Ng:2010jg}. This effect could lead to a long-range ordered hydrogenation of the graphene/Ni(111) layer. Two overstructures would be plausible in this picture, a trigonal ($\sqrt{3}$x$\sqrt{3}$)\textbf{R}30$^\circ$ structure and a rectangular $\left( \begin{smallmatrix} 2&1\\ 0&2 \end{smallmatrix} \right)$ structure: the latter is close to what we observe here.
Fig. \ref{fig:illustration} shows an illustration of this rectangular structure (green dots denote the positions of the H atoms), which matches relatively well our observations (green rectangle). For comparison, the graphene unit cell is illustrated by an orange rhombus.
Our findings predict a structure that is 14~\% longer in the $\overline{\Gamma \mathrm{M}}$ direction and 5~\% shorter in the $\overline{\Gamma \mathrm{K}}$ direction, compared to the positions of the corresponding C atoms on a pristine graphene/Ni(111) layer. 
\textit{Ab initio} calculations of the hydrogenation of a free-standing graphene lattice predict a stretching of the C--C bonds by about 5--8~\% due to a change from sp$^2$ to sp$^3$ bonding \cite{Sofo:2007bp,Leenaerts:2010dw}. A change in bond length should thus not be sufficient to explain the observed unit cell length of the overstructure. However, considering the underlying geometry of the graphene lattice, we find the proposed structure to be the only plausible explanation. 

The rotation of the observed superstructure by 4$^\circ$ appears somewhat puzzling at first glance. However, rotated graphene structures have also been observed in bilayer graphene \cite{Zhao:2011hw} and rotations of superstructures along non-symmetry angles have been observed, e.g., for physisorbed noble gas layers on graphite \cite{Shaw:1978uv,Fain:1980gt,Calisti:1982uo} and for chemisorbed alkali layers on metal substrates \cite{Doering:1983tx,Doering:1984wf,Aruga:1984ut}. Moreover, graphene grown above 500~$^\circ$C by chemical vapour deposition on the Ni(111) surface exhibits also rotated domains \cite{Patera:2013dm}, and since the adsorption energy in the eV range of an H atom is considerably larger than the calculated adsorption energy per C atom in the graphene layer of about 0.16~eV \cite{Bianchini:2014fa}, the hydrogenation may cause a reorientation of the graphene layer.

\textit{Ab inito} calculations would be a helpful tool to test the proposed structure; yet the size of the resulting supercell required for a periodic DFT calculation would be at the border of what is possible in terms of computational feasibility at present. Other possibilities include running a calculation for the hydrogenated graphene modelled as a non-periodic graphene island. However, the dimension of the island and the consequent overall offset with the Ni substrate would inevitably be arbitrary and it might be difficult to obtain completely reliable results. 

Our proposed rectangular overstructure would correspond to a theoretical H/C ratio of 12.5~\%. This is much lower than the 20~\% deduced from the attenuation of the specular reflection. Since diffraction from the rectangular structure is much weaker than diffraction found at the position of the graphene peaks, it seems reasonable to assume that a substantial part of the adsorbed hydrogen is either arranged in a graphone-like geometry or in a disordered manner and is therefore not observable for us via diffraction. (Such a graphone-like structure has been observed by STM e.g. for graphene on copper \cite{Lin:2015gq}.)
This would also explain why the observed H/C ratio is higher than the theoretical ratio of the rectangular structure. 
An H/C ratio of 16~\% has been previously observed on GR/Ni(111), which corresponds to a (trigonal) ($\sqrt{3}$x$\sqrt{3}$)\textbf{R}30$^\circ$ superstructure \cite{Ng:2010jg} whereas we did not observe such a trigonal structure. The ratio observed in our study is also much lower than the 50~\% ratio that was found in an X-ray photoelectron spectroscopy/TPD study of the same system \cite{Zhao:2015fu}. However, in the latter publication, hydrogenation was performed at low temperature (170~K), while the low coverage saturation was observed after hydrogenation at room temperature. 

Unfortunately, we did not perform adsorption or diffraction measurements at such low temperatures since we did not expect to observe any hydrogenation within reasonable times at such low temperatures (extrapolating the signal decay rate to 170~K predicts a characteristic time of adsorption of $\simeq 4.1\cdot10^{4}$~s). The overall picture is thus that of a hydrogenation that depends on both, surface temperature and coverage, where large overstructures appear at room temperature and low coverage, while at high coverage, a denser structure is likely to form. 

\subsection{Dynamics}

Helium spin-echo studies reveal information about surface diffusion. One of the aims of this study was to gain information about a possible diffusion of the adsorbed H atoms on the graphene lattice. Since the accessible spin-echo time range of the HeSE apparatus is restricted to about 1.2~ns, a time window that is considerably shorter than the expected time scale of H atom diffusion, we can only extrapolate the available data to set a lower limit on the characteristic time $\tau$ between jumps of a hydrogen atom.

Spin-echo measurements were performed at 400~K, 500~K and 600~K, respectively, at a momentum transfer $\Delta K = 1$~\AA$^{-1}$. At 500~K and 600~K, a constant overpressure of atomic hydrogen was applied in order to ensure a constant surface coverage of about 0.08~ML during the spin-echo scans, which was deduced by monitoring the specularly reflected scattering signal. 

The jump diffusion process is described by the Chudley-Elliott model \cite{Chudley:1961tt}: The incoherent ISF obtains the form of an exponential decay:
\begin{equation}
	I(\Delta\mathbf{K},t)_{inc} = I_{inc}(\Delta\mathbf{K},0) \exp[-\Gamma(\Delta\mathbf{K})t],
\end{equation}
with the decay rate 
\begin{equation}
	\Gamma(\Delta\mathbf{K})= \frac{1}{n\tau}\sum_{\mathbf{l}_j}\left[1-\exp(-i\Delta\mathbf{K}\cdot\mathbf{l}_j)\right].
\end{equation}
Here, $\mathbf{l}_j$ is the vector that connects beginning and end of the $j$th possible jump. In the geometry of jumps between neighbouring on-top adsorption sites, along the $\overline{\Gamma \mathrm{M}}$ direction the decay rate can be approximately given by the form for a hexagonal Bravais lattice:
\begin{equation}
\Gamma(\Delta K)= \frac{1}{3\tau}\left[\sin^2(\Delta K\,l/2)+2\sin^2(\Delta K\,l /4)\right].
\label{eq:gamma}
\end{equation}

For stability reasons we fitted a linearised model function to the data, using a linear least squares fitting algorithm.

From this we obtained lower bounds for the fitting parameter  $\tau$, which is the the inverse jump rate, or characteristic time between two jumps of a hydrogen atom. The lower bounds for $\tau$ are listed in Tab. \ref{tab:res}, within the corresponding one-sided confidence intervals of 1 $\sigma$ (84.2~\%) and 2 $\sigma$ (97.8~\%). While at 400~K, we could deduce a lower bound for $\tau$ of 75~ns, at 600~K this lower limit is at 4~ns since with increasing temperature the polarisation amplitude experiences a stronger initial drop due to surface phonon processes. We also list the inverse jump rates predicted by path integral molecular dynamics (PIMD) calculations by Herrero \textit{et al.} \cite{Herrero:2009vn} for comparison. Our experimental findings are not sufficient to draw further conclusions since the predicted jump times are about two orders of magnitude longer than the observed lower limits, but our findings are in line with present theory.

\begin{table}[htb]
	\centering
	\begin{tabular}{l|rrr}
	Temperature & 400~K & 500~K & 600~K \\
	\hline
	\hline
	\noalign{\vskip 2mm} 
  Prediction according to \cite{Herrero:2009vn} & 73000~ns & 1300~ns & 93~ns\\
	84.2~\% confidence & 1000~ns & 200~ns & 6~ns\\
	97.8~\% confidence & 75~ns & 45~ns & 4~ns \\
	\hline
\end{tabular}
\caption{Lower limits for the characteristic time $\tau$ between jumps at different temperatures, demanding statistical confidences of 1 and 2 $\sigma$. Also listed are the values for $\tau$ predicted by PIMD calculations by Herrero \textit{et al.} \cite{Herrero:2009vn}.}
\label{tab:res}
\end{table}

\section{Summary and Conclusion}
\label{sec:conclusion}

Graphene/Ni(111) was hydrogenated under ultra high vacuum conditions and studied by helium-3 diffraction and spin-echo spectroscopy. By measuring isothermal sorption rates, activation barriers for adsorption, $E_a=(89\pm7)$~meV, and for desorption, $E_d=(1.8\pm0.2)$~eV, were found. Diffraction scans revealed an ordered hydrogenation of the graphene surface creating a 4$^\circ$ rotated rectangular structure. Spin-echo studies allowed us to set lower limits on the jump times $\tau$ for hydrogen diffusion at temperatures in the range 400--600~K which are in accordance with PIMD calculations \cite{Herrero:2009vn}.

\section*{Acknowledgements}
\label{sec:acknowledgements}
The authors would like to thank R. Weatherup and D. Far\'ias for their advice on the graphene growth procedure as well as M. Sacchi for many useful discussions regarding DFT calculations.
	
This work is part of the Ph.D. project of E. Bahn who would like to thank the Ecole Doctorale de Physique of the Universit\'e de Grenoble for funding. A. Tamt\"ogl acknowledges financial support provided by the FWF (Austrian Science Fund) within the project J3479-N20.

\section*{References}
\footnotesize{
	\bibliography{main.bib} 
	
	\bibliographystyle{model3-num-names}
}

\end{document}